# Performance Evaluation of Organic Emulsion Liquid Membrane on Phenol Removal

**Authors and Affiliation**

Y. S. Ng [a], N. S. Jayakumar [a], M. A. Hashim [a]*

[a] Department of Chemical Engineering, University of Malaya, 50603 Kuala Lumpur, Malaysia

*Corresponding author*

| | |
|---|---|
| Email | : alihashim@um.edu.my |
| Telephone number | : +603-79675296 |
| Fax number | : +603-79675319 |
| Postal Address | : Department of Chemical Engineering, Faculty of Engineering, University of Malaya, 50603 Kuala Lumpur, Malaysia |

**Abstract**

The percentage removal of phenol from aqueous solution by emulsion liquid membrane and emulsion leakage was investigated experimentally for various parameters such as membrane: internal phase ratio, membrane: external phase ratio, emulsification speed, emulsification time, carrier concentration, surfactant concentration and internal agent concentration. These parameters strongly influence the percentage removal of phenol and emulsion leakage. Under optimum membrane properties, the percentage removal of phenol was as high as 98.33%, with emulsion leakage of 1.25%. It was also found that the necessity of carrier for enhancing phenol removal was strongly dependent on the internal agent concentration.

# 1. Introduction

Phenol is a toxic substance which is normally present in wastewater generated from refineries, pharmaceutical and petrochemical operations [1, 2] and even in small quantities, it is toxic to living organisms. According to the Health and Safety Guide No. 88, 1994 [3], phenol is corrosive to the skin and eyes and it is readily absorbed by the living tissues in liver and lung. It can cause gastrointestinal irritation, tissue erosion, protein degeneration, systemic effect such as respiratory distress, methaemoglobinaemia, neurological effect and finally death [1-3]. It has an unpleasant smell that irritates the respiratory tract even in dilute concentration. Inhalation of phenol can cause anorexia, weight loss, headache, other symptoms [3] and untreated phenol-containing wastewater will pollute the water environment and harm aquatic ecosystems. Based on the Health and Safety Guide No. 88, 1994 [3], the lethal dose $LD_{50}$ for fish and crustacean is around 7mg/L. The Malaysia Environmental Quality Act 1974 states that the maximum phenol concentration for discharged effluent is 1ppm, while phenol concentration from industrial processing effluent can be in range of 2.8-6900 ppm [2]. Therefore, it is necessary to remove the phenol from industrial effluents before they are discharged into the water stream.

There are a variety of treatment methods that have been applied for phenol removal. Popular amongst these are activated carbon adsorption [1, 4-6], chemical oxidation [7], liquid membrane [8-14] and biological treatment [15, 16].

In comparison to liquid membrane, biological treatment is not normally suitable for wastewater with high phenol concentration such as that from refinery, petrochemical and pharmaceutical operations [17]. Chemical oxidation requires a large amount of oxidizing agent under high operating conditions [2] and with a risk of incomplete oxidation and result in a more toxic product [7]. Meanwhile, activated carbon adsorption can effectively remove organic compounds such as phenol [1, 4-6] but this method has a drawback in that the activated carbon is expensive and difficult to regenerate due to chemisorption of phenol and the degradation of carbon [2].

Liquid membranes have shown potential for the removal of phenol from wastewater. They are selective permeable materials that transport certain targeted solutes. Among the different types of liquid membranes, emulsion liquid membrane (ELM) provides several advantages such as a high interfacial area for extraction, versatility, relatively low cost and a non-dependence on equilibrium consideration [18]. Li [19] was the first to introduce ELM with the purpose of increasing the interfacial area to shorten the diffusion path.

In the application for wastewater treatment, ELM consists of water-oil-water system whereby the oil phase (membrane) acts as a selective barrier and trapped the aqueous stripping agent (internal phase) inside them. The emulsion will then disperse into the wastewater (external phase) for the extraction of the targeted solutes. The solutes are then transferred from the external phase into the membrane and are then stripped down by the internal phase, the degree of which is dependent on the concentration gradient between the two phases. The internal phase will repress the activity of the targeted solutes so as to maintain the concentration gradient between the external phase and the internal phase [2]. ELM has been applied for metals and organic recovery [8] but wider application is limited due to emulsion instability and swelling.

With the objective of increasing emulsion stability, Park and Chung [9] and Mortaheb *et al.* [14] tested different surfactants on ELM and it was found that the emulsion stability can be enhanced by increasing the surfactant concentration. Other researchers [9, 13, 14, 20] have reported their works on the enhancement of emulsion stability and efficiency through the addition of stabilizer such as PIB, different emulsification methods and vortex column operation. In addition to the parameter mentioned, there are several other parameters that can influence emulsion stability such as volume ratio, temperature, stirring speed, internal stripping agent concentration, surfactant concentration, pH [9, 12-14, 21-23]. In a recent work involving a carrier Cyanex 923, Cichy *et al.* [10] and Reis *et al.* [11] found that the formation of carrier-phenol complexes in the wastewater enhance phenol extraction and recovery. However, Studies on the interrelationships between the carrier concentration and the above mentioned parameters for ELM are limited.

Thus, the emulsion stability and efficiency can be assumed to be controlled by the inherent membrane properties and the operating parameters. Example of the former are membrane: internal phase ratio, membrane: external phase ratio, emulsification time, emulsification speed, carrier concentration, surfactant concentration and internal agent concentration and the latter are pH of external phase, stirring speed and the stirring mechanism. The objective of the present work is to evaluate the effect of membrane properties on the ELM efficiency and emulsion leakage as well as the interrelationship between the internal stripping agent concentration and the carrier concentration.

## 2. Experimental

*2.1 Materials*

The phenol crystals, NaOH solid and Span 80 were supplied by Merck while Cyanex 923 as a carrier was bought from CYTEC and kerosene was supplied by ACROS Organics. Spectrophotometer SECOMAM-XT5-XTD was used for measuring absorbance while ULTRA TURRAX IKA-T25 was used as a high speed homogenizer for emulsion preparation and IKA Lab-Egg Overhead Stirrer was used as an ELM stirrer.

*2.2 Preparation of Phenol Solution and NaOH Solution*

Phenol solution of 300ppm was prepared by dissolving phenol crystals in distilled water. NaOH solution was prepared by the same method as phenol solution preparation, i.e. dissolving NaOH solids into distilled water.

*2.3 Preparation of Emulsion Liquid Membrane*

For emulsion liquid membrane (ELM) preparation, several publications [9, 11, 14, 19, 20, 22, 23] were referred to. Span 80 was used as a surfactant for ELM due to its popularity as an emulsifier for liquid membrane while Cyanex 923 was proposed to be the carrier for extraction of phenol [10, 11, 24, 25]. Kerosene was used as a diluent for the membrane [9, 21]. For the initial experiments, a surfactant: carrier: diluent ratio of 2:2:96 was used. An emulsion of volume 12mL was prepared by mixing the surfactant, carrier and diluent in a beaker together with 0.5M NaOH solution as an internal stripping agent (internal phase) in a ratio of 1:1 by volume. The mixture of W/O was then emulsified by using a high speed homogenizer ULTRA TURRAX IKA-T25, operating at a rotational speed of 8000rpm for 3 minutes so as to obtain a milky white colour liquid membrane. The parameters such as membrane: internal phase ratio, membrane: external phase ratio, emulsification time, emulsification speed, carrier concentration, surfactant concentration and internal agent concentration were varied so as to observe their effects on the percentage removal of phenol and emulsion leakage.

*2.4 Experiment of Phenol Treatment*

Calibration curve for absorbance-phenol and phenolate concentration were prepared for checking the absorbance of phenol solution by using different known concentration samples. The ELM prepared was dispersed into phenol aqueous solution (external phase) in a beaker in a ratio of membrane: external phase as 1:2 by volume. The mixture was stirred by IKA Lab-Egg overhead stirrer with a low rotational speed of 400rpm for 4 minutes. A 1mL of phenol aqueous sample was taken and analyzed by UV-Vis Spectrophotometer SECOMAM-XT5-XTD for phenol concentration. Detection of phenol and sodium phenolate can be observed at an absorbance value of 270nm [26] and 290nm [27], respectively. The concentration of phenol and phenolate were estimated from the absorbance-phenol/phenolate concentration calibration curves. The percentage removal of phenol was then determined by Equation (1):

$$\text{Percentage removal of phenol} = \frac{\text{Initial concentration - concentration of samples}}{\text{Initial concentration}} \times 100 \quad (1)$$

*2.5 Detection of Emulsion Leakage*

Emulsion leakage can be justified by measuring the presence of sodium phenolate in the analyzed samples. This value was then compared with the total amount of sodium phenolate that was produced based on the amount of phenol removed using mass balance. Phenol permeated into the liquid membrane and reacted with NaOH, which was the internal stripping agent to yield sodium phenolate and water. The reaction is shown as Equation (2):

$$C_6H_5OH + NaOH \rightarrow C_6H_5ONa + H_2O \tag{2}$$

Sodium phenolate cannot diffuse back into the external phase through liquid membrane [9] due to the selectivity of the membrane. Hence, it was not detected in the external phase, which in this case, was phenol aqueous solution. The presence of sodium phenolate in the phenol aqueous solution indicated emulsion leakage in the system, and this was estimated by Equation (3):

$$\text{Emulsion leakage, \%} = \frac{\text{Sodium phenolate concentration in samples}}{\text{Theoretical sodium phenolate concentration (fully leakage)}} \times 100 \tag{3}$$

## 3. Results and Discussions

*3.1. Phenol Removal Efficiency*

The experiments were carried out in duplicate and the results obtained were within 2% deviation. The percentage removal of phenol and the emulsion leakage through varying the experimental parameters are as shown in Table 1.

*3.1.1 Membrane: Internal Phase Ratio*

The percentage removal of phenol increases with the increment of the membrane: internal phase ratio. As illustrated in Table 1, an optimum removal of over 90% was obtained at the ratio of 3:1 by volume and above while emulsion leakage was found to be reduced as the membrane: internal phase ratio was increased. As membrane: internal phase ratio increased, more stable emulsion droplets can be formed by an increment of the membrane phase to encapsulate the internal agent. The strength of the emulsion against leakage was increased under a high ratio [14]. However, it was also found that increasing the membrane: internal phase ratio beyond 3:1 did not enhance phenol removal. This phenomenon could be due to the built-up resistance around the membrane at the high membrane: internal phase ratio. The increase in thickness of the membrane offered resistance that slowed down the phenol permeation rate. In term of emulsion leakage, an increase of membrane: internal phase ratio from 3:1 to 5:1 reduced the emulsion leakage significantly, from 14.50% to 3.51%. In this study, membrane: internal phase ratio of 5:1 can be considered as the optimum ratio since it has lowest emulsion leakage.

Table 1: Effect of different parameters on the percentage removal of phenol and emulsion leakage at 25°C.

| Parameters | Variables | Percentage Removal of Phenol, % | Emulsion Leakage, % |
|---|---|---|---|
| Membrane: Internal Phase Ratio | 1:1 | 54.57 | 50.83 |
|  | 2:1 | 83.44 | 27.12 |
|  | 3:1 | 93.76 | 14.90 |
|  | 4:1 | 93.77 | 8.25 |
|  | **5:1** | **93.80** | **3.51** |
| Membrane: External Phase Ratio | 1:1 | 90.79 | 6.67 |
|  | **1:2** | **93.80** | **3.51** |
|  | 1:3 | 89.01 | 4.74 |
|  | 1:4 | 85.16 | 4.96 |
| Emulsification Speed (rpm) | 4000 | 84.46 | 40.45 |
|  | 5000 | 86.79 | 34.71 |
|  | 6000 | 91.36 | 9.93 |
|  | 7000 | 91.80 | 8.51 |
|  | **8000** | **93.80** | **3.51** |
| Surfactant Concentration, %(v/v) | 1 | 93.16 | 4.35 |
|  | **2** | **93.80** | **3.51** |
|  | 3 | 91.84 | 3.56 |
|  | 4 | 85.94 | 4.19 |
| Emulsification Time (min) | 1 | 81.05 | 9.59 |
|  | 2 | 89.00 | 5.93 |
|  | 3 | 93.80 | 3.51 |
|  | **5** | **96.94** | **1.95** |
|  | 7 | 90.91 | 4.03 |
|  | 10 | 82.37 | 10.91 |
| Carrier Concentration, %(v/v) | **0** | **98.33** | **1.25** |
|  | 1 | 97.61 | 1.40 |
|  | 2 | 96.94 | 1.95 |
|  | 3 | 81.49 | 6.70 |
| Internal Agent Concentration (M) | 0.02 | 33.56 | 1.60 |
|  | 0.06 | 57.98 | 1.03 |
|  | **0.50** | **98.33** | **1.25** |
|  | 0.75 | 98.61 | 1.70 |
|  | 1.00 | 97.77 | 1.95 |

### 3.1.2 Membrane: External Phase Ratio

The membrane: external phase ratio was another parameter that affected the performance of the ELM. From Table 1, an optimum ratio of membrane: external phase can be taken as 1:2 by volume, as this ratio resulted in the highest percentage removal of phenol and the lowest emulsion leakage. The study revealed that with the increase of the external phase volume, the membrane area per total external volume in the system was being reduced [23]. This may lead to the reduction of phenol permeation flux into the membrane phase. However, the experiments proved that the percentage removal of phenol was lower with more emulsion leakage at a low membrane: external phase ratio of 1:1. This trend was observed by Lin *et al.* [21] and the explanation given was that the coalescence of emulsion occurred under a high membrane: external ratio due to the ineffectiveness of dispersion by stirring. The coalescence of emulsion reduces the total surface area for extraction, thus reducing the percentage removal of phenol. In other ratio of 1:2, 1:3 and 1:4, the emulsion leakage is not much affected by membrane: external phase ratio.

*3.1.3 Emulsification Speed*

Table 1 illustrates that with an increase in the emulsification speed, the percentage removal of phenol increases while the emulsion leakage reduces. Higher emulsification speed yielded smaller size of emulsion liquid droplets which make the emulsion more efficient to extract phenol from the external aqueous phase. According to Djenouhat *et al*. [20] and Gasser *et al*. [22], smaller size of emulsion liquid droplets gave better dispersion in the external phase while providing more interfacial surface area for mass transfer. They also suggested that high emulsification speed gave good dispersion of internal phase in the membrane phase, thus providing better emulsion stability with lower leakage due to the higher coalescence time. In this study, the optimum emulsification speed in this study was found to be 8000rpm.

*3.1.4 Surfactant Concentration*

A surfactant concentration of 2% volume caused the highest percentage removal of phenol and the lowest emulsion leakage, as shown in Table 1. Surfactant was added as an emulsifier for the liquid membrane and it acted as a protective barrier between the external phase and the internal phase, preventing emulsion leakage. An increment of the surfactant concentration will increase the stability of the liquid membrane and hence reduce the emulsion leakage, as reported by several researchers [9, 13, 22]. The increment of surfactant concentration lowered the membrane's surface tension [21] and yielded smaller globules which led to a higher contact area [14]. However, an excess of surfactant concentration caused a low percentage removal of phenol in this study. This may be due to the increment of membrane viscosity and thickness of the membrane [9, 14] under high surfactant concentration. In these experiments, the increment of surfactant concentration beyond its optimum concentration of 2% led to a higher resistance for phenol permeation into the internal phase, which consequently reduced the percentage removal of phenol. However, the effect of surfactant concentration on emulsion leakage was not obvious.

*3.1.5 Emulsification Time*

Table 1 also illustrated that emulsification time of 5 minutes yield the highest percentage removal of phenol and the lowest emulsion leakage. An insufficient emulsification time caused significant emulsion leakage due to the coalescence of larger globules in a shorter time, as was observed in this study for emulsification time of 1-3 minutes. According to Gasser *et al*. [22], a low emulsification time will cause the formation of large globules, where a less interfacial area reduced the mass transfer rate. On the other hand, a prolonged emulsification time of above 5 minutes led to a lower percentage removal of phenol and a higher emulsion leakage in this study. This was probably caused by the coalescence of the internal droplets [20] and the high shearing rate [22].

*3.1.6 Carrier Concentration*

The amount of carrier present in the membrane affected the performance of the ELM significantly. Theoretically, Cyanex 923 enhanced extraction rate of phenol by forming complexes with phenol which were permeable through the solvent/liquid membrane [10, 11, 24, 25]. However, the carrier also tended to change membrane properties [20] and formed a reversed emulsion by swelling [22], which led to the rupture of emulsion, causing instability and leakage of the emulsion. In these experiments, this effect was not well observed in 0-2% carrier concentration, as shown in Table 1. However, in carrier concentration of 3%, it was found that the carrier caused more membrane swelling rather than enhancing the phenol removal rate since more emulsion leakage was detected and the reduction of percentage removal of phenol was observed from >96% to 81.49%. The experiments also showed that when the carrier concentration was less than 2%, the carrier has no significant effect on the percentage removal of phenol since the result was within the experimental error range, and the emulsion leakage was under 2%. This may be due to the extraction rate enhancement by Cyanex 923 is insignificant under ELM process which has a thin diffusion path. High distribution coefficient of phenol in membrane phase is not necessary. Thus, the increment of extraction rate is not really an advantage in comparison to the instability given by the presence of Cyanex 923. In this study, the optimum carrier concentration can be taken as 0%. These findings are in line with that of Frankenfeld and Li [18], where it is reported that the phenol transport in liquid membrane is a passive transport and no carrier is necessary.

*3.1.7 Internal agent concentration*

The data in Table 1 also shows that the best percentage removal of phenol with low emulsion leakage can be achieved by using 0.5M NaOH. As the NaOH concentration was increased from 0.02M -0.5M, more phenol can be stripped down from the membrane phase due to the higher contact rate between phenol and NaOH in the internal phase. At low NaOH concentration, there was insufficient NaOH to remove the phenol from the membrane phase. Stripping process was slowed down and the saturation of phenol on membrane occurred. An increment of NaOH concentration can enhance the stripping process of phenol from the membrane phase. However, as the NaOH concentration increased beyond optimum concentration of 0.5M, the percentage removal of phenol was not enhanced. As for emulsion leakage, Mortaheb *et al.* [14] stated a high NaOH concentration will increase the pH of internal phase and osmotic swelling may occur under a high pH difference between the external phase and the internal phase. Several studies also showed that the difference in ionic strength between two phases led to the transportation of water from the external phase to the internal phase and an increase of the internal emulsion volume which further causing the emulsion leakage [11, 21]. However, in the experiments, this effect was not observed as the emulsion leakage was maintained within 2%. This may be due to the thickness of the membrane capable of maintaining the emulsion stability under high membrane: internal phase ratio. Considering the percentage removal of phenol, 0.5M NaOH was chosen as the optimum value in this experiment.

*3.2 Relationship between Carrier Concentration and Internal Agent Concentration*

A set of experiments were carried out to study the interrelationship between the carrier concentration and the internal agent concentration and its influence on the efficiency of the ELM. Fig. 1 shows that at low NaOH concentration, the presence of the carrier enhances phenol removal while at high NaOH concentration, the presence of carrier does not have any significant advantages. This may be due to the driving force difference that was exerted between NaOH and phenol. In addition, the solubility of phenol in kerosene may be a factor causing this phenomenon.

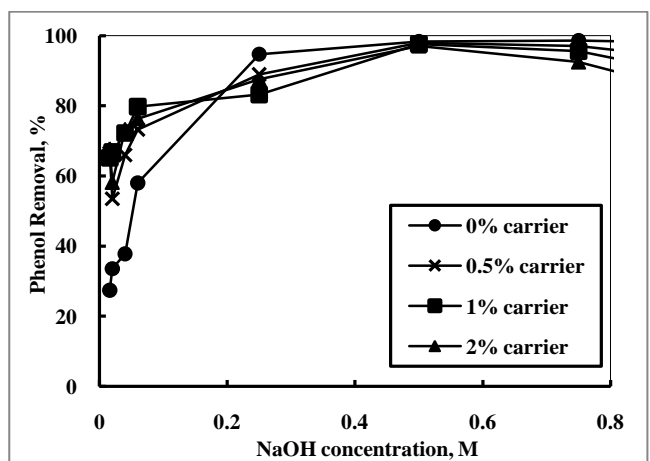

Fig. 1: Comparison of carrier concentration and NaOH concentration's effect on the percentage removal of phenol

Through a simple solvent extraction experiment, kerosene was found to dissolve a small amount of phenol from the phenol aqueous solution. In 0% carrier concentration, the phenol dissolved by the kerosene was stripped down by the NaOH in the emulsion, following the mechanism suggested by Frankenfeld and Li [18] for passive transport of organic compounds. Fig. 2 divides the passive transport mechanism of phenol in ELM into 3 sections: Phenol dissolves into the membrane phase due to the concentration gradient between the external phase and kerosene/membrane phase (Fig. 2-a). The transport of phenol in kerosene occurred due to another concentration gradient between the membrane phase and the internal phase (Fig. 2-b). The driving force enables the diffusion of phenol to the internal phase and finally stripped down by NaOH (Fig. 2-c). Due to the phenol concentration gradient exerted between the kerosene and both the external phase and the internal phase, phenol continued to dissolve in kerosene and stripped by NaOH. A continuous process was formed until equilibrium was reached. Thus, in low NaOH concentration, the stripping rate of phenol was the limiting step in the system. Fig. 1 confirms that the percentage removal of phenol is low in low NaOH concentration region. This could be caused by saturation of phenol in kerosene as a result of low stripping rate.

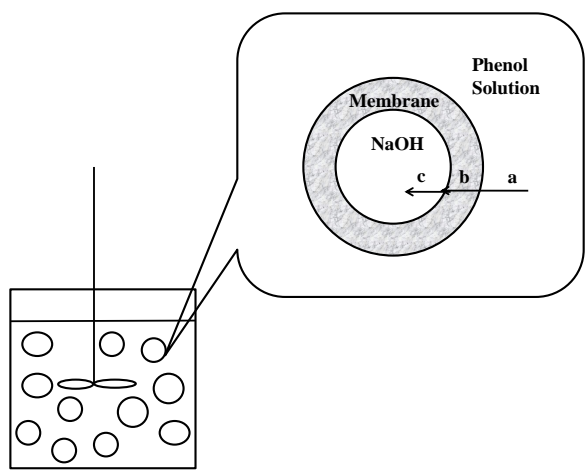

Fig. 2: Passive transport mechanism of phenol in ELM

The result in Fig. 1 also shows that in the presence of carrier, percentages removal of phenol was enhanced at low NaOH concentration. However, the increment of carrier concentration from 0.5% to 2% has less effect on further enhancement of phenol removal when NaOH concentration is <0.5M. It was also found that the percentage removal of phenol was decreased under 0.75M NaOH as increased of carrier concentration. In the presence of carrier, phenol formed complexes with Cyanex 923. The complexes were easily dissolved in kerosene, hence the mass transfer rate of phenol from external phase to the membrane phase was enhanced [2, 11]. More phenol complexes were available in the membrane. The effective collision rate between NaOH and phenol complexes increased and served as a major driving force during low NaOH concentration condition. Hence phenol can be removed rapidly even at low NaOH concentration in comparison to the 0% carrier concentration. However, in high NaOH concentration, the driving force served by Cyanex 923 was insignificant compared to the driving force that was exerted by high NaOH concentration. Even though Cyanex 923 caused more phenol to be dissolved in kerosene, it contributed to the instability of emulsion as well [22], as discussed in Section 3.1.6. The instability of emulsion caused leakage of sodium phenolate and NaOH to the external phase when the water transport was occurred under high NaOH concentration, as stated in Section 3.1.7. This consequently traded off the advantage provided by the carrier and reduced the percentage removal of phenol. Overall, in comparison to 0% carrier concentration under high NaOH concentration, the phenol removal efficiency was lower in the presence of carrier.

*3.3 Effect of Time*

Utilising the parameters for optimum membrane properties obtained from Section 3.1, a higher volume experiment involving an ELM of 120mL, was undertaken. This volume was mixed with 240mL of 300ppm phenol solution and stirred using IKA lab-Egg Overhead Stirrer. The parameters employed are as listed in Table 2.

Table 2: Parameters for optimum membrane properties for emulsion liquid membrane

| Parameters | Value |
| --- | --- |
| Membrane: Internal Phase Ratio | 5:1 by volume |
| Membrane: External Phase Ratio | 1:2 by volume |
| Emulsification Speed (rpm) | 8000rpm |
| Surfactant Concentration, %(v/v) | 2% by membrane volume |
| Emulsification Time (min) | 5 minutes |
| Carrier Concentration, %(v/v) | 0% |
| Internal Agent Concentration (M) | 0.5M |

Samples having volume of 1mL of samples was taken for analysis at different time intervals. Fig. 3 illustrates the plot of percentage removal of phenol versus time. Phenol was removed rapidly once it came in contact with the ELM. The percentage removal of greater than 90% was achieved within one minute. The percentage removal of phenol increased slowly until a maximum of 98.46% was achieved within 4 minutes. It is to be noted that the percentage removal of phenol is almost similar to the small-scale experiment (Section 3.1), thus demonstrating that the parameters for optimum membrane properties can be applied for different scales. Therefore, an effective treatment of phenol with low emulsion leakage can be achieved by utilising a combination of parameters for optimum membrane properties.

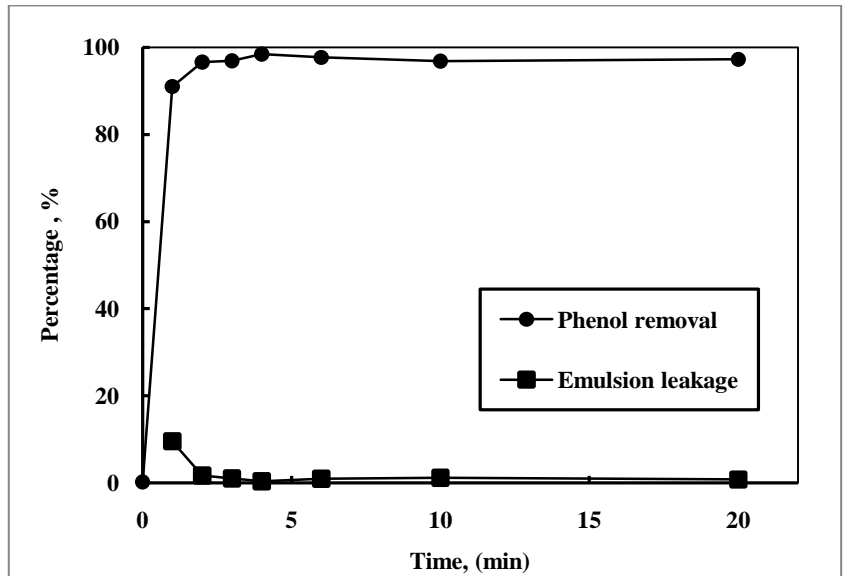

Fig. 3: Percentage removal of phenol versus time

*3.4 Comparison of Parameters for Optimum Membrane Properties*

Table 3 provides a comparison of ELM parameters for optimum membrane properties for phenol removal between the present work and that of other researchers. It shows that, in general, the percentage removal of phenol can be maintained at >95% by altering the parameters for membrane properties. The present study requires a higher membrane: external phase ratio and a higher membrane: internal phase ratio. But, with reference to previous works [9, 11-14, 21], this ratio can be reduced, without affecting the emulsion efficiency by altering other parameters such as the types of surfactant and its concentration, emulsification speed and time and internal agent concentration. In fact, the emulsion efficiency can be further enhanced through a detailed study on the interrelationships between the various parameters such that it can be widely applied in industries.

Table 3: Comparison of parameters for optimum membrane properties between present work and other studies for phenol removal.

| | **Present Work** | [9] | [11] | [12] | [13] | [14] | [21] |
|---|---|---|---|---|---|---|---|
| Diluent | Kerosene | Kerosene | ShellSol T | ShellSol T | Soltrol 220 | Petroleum solvent | Kerosene |
| Carrier | Cyanex 923 | - | Cyanex 923 | - | - | - | - |
| Surfactant | SPAN 80 | SO-10, Arlacel83 | Polyamine ECA4360 | Polyamine ECA4360 | SPAN 80 | synthesize surfactant | SPAN 80 |
| Membrane: Internal Ratio | 5:1 | 1:1 | 2:1 | 10:x | 0.46:1 | 2:1 | - |
| Membrane: External Ratio | 1:2 | 1:5 | 1:10 | x:1 | 1:3 | 1:10 | 1:10 |
| Emulsification Speed | 8000rpm | 1200rpm | 7000rpm | 1000rpm | - | 15000rpm | 4000rpm |
| Surfactant Concentration | 2% v/v | SO-10 5% wt, Arlacel83 7% wt | 2% wt | 2% wt | 5% w/v | 3% wt | 5% |
| Emulsification Time | 5 min | 10 min | 15 min | 15 min | - | 20 min | 20 min |
| Carrier Concentration | 0% | - | 2% wt | - | - | - | - |
| Internal Agent Concentration | 0.5M | 3% wt | 0.5M | 0.5M | 0.5N | 1% wt | 0.5% wt |
| Extraction Time | 4 min | 10 min | 3-6 min | 2-10 min | 10min | 4 min | |
| Extraction Efficiency | 98.33% | 99.55% | 98% | >99% | 96.18% | >95% | 98% |
| Emulsion Leakage | 1.25% | <1.5% | 1.2 | - | - | <1% | - |

## 4. Conclusions

The study on the percentage removal of phenol and emulsion leakage for ELM was conducted experimentally and the following conclusions can be made:

1. Emulsion efficiency is strongly influenced by membrane: internal phase ratio, membrane: external phase ratio, emulsification speed, emulsification time, carrier concentration, surfactant concentration and internal agent concentration. Optimum membrane properties can yield a stable ELM with a good percentage removal of phenol of up to 98.33% and the emulsion leakage can be maintained at 1.25%.

2. Internal agent concentration has a significant effect on dictating the necessity of carrier for enhancing phenol removal. The carrier has a significant role in increasing the ELM's efficiency when the internal stripping agent, NaOH concentration is less than 0.2M in this study. The carrier concentration has less effect on percentage removal of phenol at low NaOH concentration. The percentage removal of phenol was decreased in high NaOH concentration as the increment of carrier concentration

3. The parameters for optimum membrane properties of ELM can be applied for different scales.

**Acknowledgement**

This work has been supported through a grant from UMRG RG033/09SUS.